\documentclass{aastex63}

\usepackage{hyperref,epsfig,graphicx,lineno,longtable}

\received{***}
\revised{***}
\accepted{***}
\submitjournal{ApJ}

\shorttitle{Downflows}
\shortauthors{Sun et al.}
\graphicspath{{./}{figures/}}

\begin{document}

\title{Fast Downflows Observed during a Polar Crown Filament Eruption}

\author[0000-0001-5657-7587]{Zheng Sun}
\affiliation{School of Earth and Space Sciences, Peking University, Beijing, 100871, China; \href{mailto:huitian@pku.edu.cn}{huitian@pku.edu.cn}}

\author[0000-0002-1369-1758]{Hui Tian}
\affiliation{School of Earth and Space Sciences, Peking University, Beijing, 100871, China; \href{mailto:huitian@pku.edu.cn}{huitian@pku.edu.cn}}
\affiliation{National Space Science Center, Chinese Academy of Sciences, Beijing 100190, China}

\author[0000-0001-6655-1743]{Ting Li}
\affiliation{National Astronomical Observatories, Chinese Academy of Sciences, Beijing 100101, China}
\affiliation{School of Astronomy and Space Science, University of Chinese Academy of Sciences, Beijing 100049, China}
\affiliation{National Space Science Center, Chinese Academy of Sciences, Beijing 100190, China}

\author[0000-0003-4618-4979]{Rui Liu}
\affiliation{CAS Key Laboratory of Geospace Environment, Department of Geophysics and Planetary Sciences University of Science and Technology of China, Hefei, 230026, China}
\affiliation{CAS Center for Excellence in Comparative Planetology, University of Science and Technology of China, Hefei, 230026, China}
\affiliation{Mengcheng National Geophysical Observatory, University of Science and Technology of China Mengcheng 233500, China}

\author[0000-0001-9491-699X]{Yadan Duan}
\affiliation{School of Earth and Space Sciences, Peking University, Beijing, 100871, China; \href{mailto:huitian@pku.edu.cn}{huitian@pku.edu.cn}}

\begin{abstract}
  
Solar filaments can undergo eruptions and result in the formation of coronal mass ejections (CMEs), which could significantly impact planetary space environments. Observations of eruptions involving polar crown filaments, situated in the polar regions of the Sun, are limited. In this study, we report a polar crown filament eruption (SOL2023-06-12), characterized by fast downflows below the filament. The downflows appear instantly after the onset of the filament eruption and persist for approximately 2 hours, exhibiting plane-of-sky (POS) velocities ranging between 92 and 144 km s$^{-1}$. They originate from the leading edge of the filament and no clear acceleration is observed. Intriguingly, these downflows appear at two distinct sites, symmetrically positioned at the opposite ends of the conjugate flare ribbons. Based on the observations, we propose that the filament might be supported by a magnetic flux rope (MFR), and these downflows possibly occur along the legs of the MFR. The downflows likely result from continuous reconnections between the MFR and the overlying magnetic field structures, and could either be reconnection outflows or redirected filament materials. We also observed horizontal drifting of the locations of downflows, which might correspond to the MFR's footpoint drifting. This type of downflows can potentially be utilized to track the footpoints of MFRs during eruptions.

\end{abstract}


\section{INTRODUCTION} \label{sec:intro}

Solar filaments are magnetic structures that confine cool plasma (with temperatures below $10^4$ K) at high densities (electron densities ranging from $10^9$ to $10^{11}$ cm$^{-3}$) within the hot solar corona (e.g., \citealt{devore2005solar,van2013structure,parenti2014solar}). We use ``filaments'' for those observed on the solar disk, where they are seen in absorption in H$\alpha$ and some extreme-ultraviolet (EUV) spectral lines. When seen above the solar limb, they appear as bright structures against the dark background, which are called ``prominences''. Throughout this paper, we will use ``filament'' to indicate the filament structure.
According to the location, filaments can be divided into active region (AR) filaments, intermediate filaments (located at the border of ARs), and quiescent filaments (located in the quiet Sun). Quiescent filaments exhibit relatively stable characteristics in contrast to other two types of filaments (e.g., \citealt{d1948comprehensive,zhou2006coronal,su2015magnetic}). Their lifetimes span from several days to several months (e.g., \citealt{isobe2006large}). They are typically within the range of  10$^4$ to 10$^5$ km in length and 10$^3$ to 10$^4$ km in thickness (e.g., \citealt{hirayama1985modern,mackay2010physics,haerendel2011droplet}).
Filaments are consistently located above the polarity inversion lines (PILs; \citealt{wang2007formation,mackay2008solar,kuckein2012active}). In the quiet Sun, however, magnetic fields are concentrated within discrete network elements that are distinctly separated from each other. The areas between these network elements exhibit significantly weak magnetic fields. Consequently, a PIL in the quiet Sun is characterized by mixed magnetic polarities, which may be identified from spatially smoothed magnetograms \citep{rondi2007photospheric,mackay2010physics}.

Filaments are known to undergo eruptions as parts of CMEs, which can significantly significantly disturb the solar-terrestrial space environment (e.g., \citealt{chen1996theory,schmieder2013solar,li2015filament,li2015high,sun2024solar}). The mechanisms of the eruptions could be classified into two categories: the magnetic reconnection and MHD instability. The former includes the tether-cutting reconnection (e.g., \citealt{moore2001onset,liu2010sigmoid,chen2018witnessing}), breakout reconnection (e.g., \citealt{shen2012sympathetic,wyper2018breakout,sun2023observation}), and flux emergence models (e.g., \citealt{wang1998filament,xu2008statistical,schmieder2013solar,duan2024formation}). The MHD instability models include the kink instability \citep{hood1979kink} and torus instability \citep{kliem2006torus}. There are often multiple mechanisms working together in filament eruptions \citep{sterling2004evidence,vemareddy2012filament,dai2022partial}. Polar crown filaments, as a special type of quiescent filament, are located at high latitudes of the Sun and exceedingly stable \citep{d1948comprehensive}. Observations of polar crown filament eruptions are very limited \citep{isobe2006large,zhou2006coronal,su2015magnetic,zou2019statistical}.

A filament is often supported by a magnetic flux rope (MFR), where the field lines are wound multiple times around the central axis (e.g., \citealt{aulanier19983,liu2020magnetic}). Observations indicate that 89\% of filaments are sustained by MFRs, with this percentage rising to 96\% for quiescent filaments \citep{ouyang2017chirality}. In three-dimensional (3D) MHD simulations, MFRs are enveloped by quasi-separatrix layers (QSLs) due to their twisted nature \citep{titov2002theory,demoulin1996quasi,demoulin1997quasi}. Typically, the footprints of QSLs wrapping the MFRs exhibit a distinctive double-J configuration, coinciding with the observed flare ribbons in ultraviolet (UV) and H$\alpha$ wavelengths \citep{demoulin1997quasi,janvier2014electric,janvier2016evolution,2015ApJ...811..139T,su2018high}. The footpoints of a MFR are thought to be situated inside the ``J" hook \citep{aulanier2012standard,janvier2013standard,jiang2018magnetohydrodynamic}. In these simulations, the double-J footprints gradually separate from each other during eruptions \citep{janvier2016evolution,jiang2018magnetohydrodynamic}. This phenomenon is directly observable in many observations, manifesting as the progressive separation of flare ribbons (e.g., \citealt{janvier2016evolution,zhao2016hooked,wang2017buildup,liu2022apparent}).

The footpoints of MFRs anchored in the double-J hooks are not fixed during eruptions. \citet{aulanier2019drifting} conducted a 3D line-tied MHD simulation to illustrate the phenomenon of footpoint drifting in MFRs during eruptions. They found that footpoints can undergo a prominent drifting due to continuous reconnection between the MFR and overlying arcades, referred to as ``ar-rf'' reconnection (``a'' for arcade field lines, ``r'' for flux-rope field lines, and ``f'' for flare loops). This has been evidenced by subsequent observations \citep{lorinvcik2019manifestations,zemanova2019observations,chen2019observational}. 
In this work, we reported a polar crown filament eruption, characterized by fast downflows near the footpoints. The downflows might be associated with the reconnection outflows in ar-rf reconnections. The observations and results are presented in Section \ref{sec:observations}. Physical mechanisms are discussed in Section \ref{sec:discussion}. Finally, we briefly summarize our results in Section \ref{sec:conclusion}.

\section{Observations and Results} \label{sec:observations}
The Atmospheric Imaging Assembly (AIA) onboard the Solar Dynamics Observatory (SDO) can provide full-disk extreme ultraviolet (EUV) images of the Sun since 2010. These images can reveal various types of atmospheric dynamics over a wide range of temperatures from $\log (T$/K) $\approx 4.0$ to $\log (T$/K) $\approx 7.0$ \citep{lemen2012atmospheric}. We analyzed 131~\AA, 171~\AA, 211~\AA,  and 304~\AA\ data with a spatial pixel size of 0.6$^{\prime\prime}$ and a temporal cadence of 12 seconds in our study. Some of the images were enhanced by using the multiscale Gaussian normalization (MGN; \citealt{morgan2014multi}) technique. In addition, the Helioseismic and Magnetic Imager (HMI) on board SDO can obtain line-of-sight (LOS) magnetograms of the entire solar disk with a spatial resolution of 0.5$^{\prime\prime}$  pixel$^{-1}$ and a cadence of 45 s \citep{scherrer2012helioseismic}.
We also used the EUV data from the Solar Ultraviolet Imager (SUVI) onboard the Geostationary Operational Environmental Satellite (GOES; \citealt{darnel2022goes}) mission. We ultilized 304~\AA\ images with a cadence of 4 minutes and a spatial pixel size of 2.5 $^{\prime\prime}$. SUVI is noteworthy for its large FOV, which can extend to $\sim$1.6 solar radii on the horizontal and $\sim$2.3 solar radii on the diagonal directions.
The Solar Upper Transition Region Imager\footnote{https://sun.bao.ac.cn/SUTRI/} (SUTRI; \citealt{bai2023solar}) also contributes to our study by providing EUV images in Ne VII 465~\AA, corresponding to a temperature of approximately 0.5 MK \citep{tian2017probing}. SUTRI's capabilities include a spatial resolution of approximately 8$^{\prime\prime}$ and a time resolution of about 30 seconds. H$\alpha$ images from the Chinese H$\alpha$ Solar Explorer\footnote{https://ssdc.nju.edu.cn/home} (CHASE; \citealt{li2022chinese}) were also used to reveal the filament morphology and dynamics. CHASE's H$\alpha$ Imaging Spectrograph (HIS) has a spectral pixel size of 0.024~\AA\ and a spatial pixel size of 0.52$^{\prime\prime}$. Using these data, we investigate a polar crown filament eruption on 2023 June 12.

Figure 1 presents the pre-eruption observations of the polar crown filament, situated between latitude 60$^{\circ}$N and 70$^{\circ}$N and spanning approximately half the solar longitude. Due to the weak magnetic field in the quiet Sun, we could not observe a clear PIL below the filament. At $\sim$02:00 UT, the filament is initiated and starts to rise, as shown in Figure 2 and the online animated version of Figure 1. Instantly after the onset, some downflows occur near the eastern footpoint of the filament, denoted by a cyan arrow in Figure 2(a). As the eruption progresses, these downflows become observable at higher altitudes, allowing us to detect similar downflows near the western footpoint, which is on the back side of the Sun (panels (c)-(d)). Panels (c)-(d) and the movie associated with Figure 1 reveal the locations of the conjugate flare ribbons (marked as blue dashed lines), moving away from each other. The white dashed lines indicate locations where downflows occur. The downflows are situated symmetrically near the opposite ends of the conjugate flare ribbons, which exhibits a double-J morphology (the blue and white dashed lines). This indicates that the locations of downflows might correspond to the hooks of the MFR \citep{titov1999basic,zhao2016hooked}. Ultilizing data with a larger FOV from SUVI, panels (e) and (f) and the related animation show the erupting filament at larger heights. These observations reveal an unwinding motion at the eastern leg of the filament, suggesting the possible presence of a twisted structure, i.e., a MFR. However, due to the weak photospheric magnetic field of the quiet Sun, it is hard to determine the exact twist number of the MFR using the technique of magnetic field extrapolation \citep{liu2016structure}. 
We chose two cuts, S1 and S2 (illustrated in Fig.2 (c)), to investigate the kinematic properties of the filament and the origin of the downflows. The corresponding space-time diagrams are shown in panels (g)-(h), with the green dashed curves indicating the leading edge of the erupting filament. The trajectory of the filament indicates a slow rise phase at a speed of $\sim$55 km s$^{-1}$ followed by a fast rise phase at a speed of $\sim$275 km s$^{-1}$ (the POS velocity calculation in this work is generally accurate to approximately $\pm$10 km s$^{-1}$), which is in accordance with previous observations of filament eruptions \citep{gosain2012multi,joshi2013study,wang2023impulsive}. The downflows are depicted as straight lines originating from the leading edge of the filament, as shown in the time-distance diagrams. The POS speed of the downflows is approximately 76 km s$^{-1}$ near the western footpoint and 100 km s$^{-1}$ near the eastern footpoint. No obvious acceleration of the downflows was observed. 

Figure 3 shows a zoom-in view of the downflows in 171~\AA, 304~\AA, and 211~\AA, within a region around the eastern footpoint. Notably, the downflows are most clearly identified in 171~\AA, while appearing less distinct in 304~\AA. The downflows manifest themselves as thread-like structures and cause EUV brightenings at lower altitudes. Moreover, the spatial distribution of these downflows reveals a dynamic pattern, characterized by a predominant westward drifting (see panels (a1)-(a5) and (b1)-(b5) and associated movie). The drifting distance reaches up to $\sim$110 Mm. We noticed that downflows near the western footpoint also exhibit drifting, as seen from the online animation related to Fig. 1. There is some eastward drifting as well, though it is not as significant in intensity as the westward drifting. To further investigate the kinematic properties of the downflows, we constructed another two time-distance diagrams, depicted in Figure 4. These diagrams highlight some distinct downflows, delineated by the blue dashed lines. Specifically, panels (a1)-(a3) show downflows occurring predominantly near cut S3 (a2) at speeds of 120-144 km s$^{-1}$, occurring around 02:05 UT. In contrast, downflows near the western edge (near cut S4, see panels (b1)-(b3)) at speeds of 92-136 km s$^{-1}$ appear approximately 25 minutes later, around 02:30 UT. Panels (b1) and (b2) reveal a notable upward trend in the origin site of these downflows over time, presumably associated with the rise of the filament. Notably, panel (b3) shows that the downflow signals in 304~\AA\ are not easy to identify after $\sim$03:10 UT, indicating a possible decrease in density of the downflows.

To investigate the impact of the downflows on the lower atmosphere, we focused on two specific regions, R1 and R2, as shown in Fig.3. We plotted the light curves of AIA 171~\AA, 211~\AA, 304~\AA, and 131~\AA\ within the selected boxes and show the results in Figure 5. The two panels reveal that the peaks of the emissions, which indicate the location of the downflows, are almost simultaneous at the four wavelengths. This suggests that the observed sudden enhancements might be caused by density enhancement or implusive heating \citep{lionello2016can}.

\section{DISCUSSION} \label{sec:discussion}

In this study, we report a polar crown filament eruption (SOL2023-06-12), characterized by fast downflows beneath the filament. The unwinding motions near the legs of the filament suggest that the filament might be supported by a MFR. The downflows appear instantly after the onset of the eruption and last for about 2 hours. The POS speeds of the downflows are in the range of 92--144 km s$^{-1}$.
These downflows appear at two distinct sites, symmetrically positioned at the opposite ends of the conjugate flare ribbons (see the blue and white dashed lines in Fig. 2). Notably, this distribution is similar to the expected double-J QSL footprints of MFRs, which suggests an association of the downflows with the hooks of the MFR. Moreover, many 3D MHD simulations have shown that the footpoints of MFRs are anchored inside the hooks of the QSL footprints \citep{aulanier2012standard,janvier2013standard}. In this way, we speculate that the downflows move along the legs of the MFR.

Many observations have shown filament materials falling back to the solar surface due to gravity in failed eruptions or partial filament eruptions (e.g., \citealt{ji2003observations,li2017plasma,kumar2023plasmoids}). In these studies, filament materials first erupt to a certain height and then fall back to the solar surface with clear acceleration. However, in our event, the downflows appear immediately as the eruption occurs (Figure 2(a) and and the online animated version of Figure 1). Additionally, we do not observe clear acceleration at the origin site of the downflows (Figure 2(g) and (h)). These observational facts suggest that the observed downflows are more likely the result of magnetic reconnection rather than filament material falling to the solar surface due to gravity. \citet{van2014coronal} reported redirected filament materials caused by reconnection between the filament and surrounding magnetic field structures during an eruption on 2011 June 7. As the filament erupted, the overlying arcades expanded laterally and reconnected with field lines of adjacent AR, forming a QSL structure within the two ARs. Some of the filament materials reached the QSL and then propagated along the newly formed inter-AR field lines to remote magnetic footpoints in the adjacent AR. \citet{liu2018disintegration} investigated a similar event, where a filament situated beneath a fan-like structure erupted. The fan-like structure redirected the filament materials to the footprints of the fan, resulting in the failure of the eruption. In both reported events, the redirected filament materials did not show clear acceleration, which is similar to our observations. Hence, we think that the downflows might be reconnection outflows or redirected filament materials. If the downflows are reconnection outflows, the initial speed of these reconnection-produced downflows should be controlled by the local Alfv$\acute{e}$n speed and subsequently influenced by gravity and drag forces in the solar atmosphere during propagation. In this event, we did not observe clear acceleration during propagation, suggesting that the gravitational force is approximately balanced by the drag force along the propagation path.

\citet{aulanier2019drifting} employed a 3D MHD simulation to investigate various reconnection processes during MFR eruptions. They identified the ar-rf reconnection mechanism, wherein the MFR reconnects with the overlying magnetic field lines. This reconnection differs from the conventional double-flare-ribbon reconnection (aa-rf), which involves the reconnection between two sides of arcade field lines. Our observed downflows appear to be produced by ar-rf reconnection, as the downflows occur near the footpoints of the erupting MFR rather than between the conjugate flare ribbons. However, the presence of two flare ribbons suggests that the aa-rf reconnection should have also happened. The absence of the discernible outflows produced in aa-rf reconnection may be explained by the weaker background magnetic field and density compared with the MFR. On the contrary, the ar-rf reconnection involving the MFR with stronger magnetic field strength and denser filament materials can produce more prominent reconnection outflows or make denser plasma redirected. \citet{aulanier2019drifting} also predicted another form of reconnection, involving two legs of the MFR reconnecting with each other, called rr-rf reconnection. A recent investigation by \citet{dudik2022filament} showed that the rr-rf reconnection can also produce prominent downflows. However, this reconnection involves both legs of a MFR approaching towards each other, which is not observed in our observations. Hence, the downflows in our event appear to be most likely associated with ar-rf reconnection. In this scenrio, as the filament-hosting MFR erupts, it continuously reconnects with the overlying magnetic field structures and produces the observed fast downflows along the legs of the MFR. 

\citet{li2016magnetic} investigated the reconnection between a filament
and surrounding coronal loops and they identified outflows near the current sheet. This type of reconnection, termed external reconnection, is similar to the reconnection observed in our event, which involves the filament reconnecting with surrounding magnetic structures. Recently, some observational and simulation works have shown that these external reconnections might cause the failure of eruptions \citep{jiang2023model,chen2023model,chen2023observations}. \citet{chen2023observations} investigated a failed filament eruption in a quadrupolar magnetic configuration with a null point. As the filament erupts and reaches the null point, it reconnects with the surrounding magnetic field lines and fails to erupt.

Another phenomenon that exhibits downward motions during solar eruptions is known as supra-arcade downflows (SADs). SADs are dark, sunward-moving voids within the supra-arcade fan, with velocities ranging from 20 to 400 km s$^{-1}$ (e.g., \citealt{samanta2021plasma,tan2023statistical}). Proposed models of SADs include cross-sections of evacuated flux tubes \citep{mckenzie2000supra}, shrinking flare loops or their wakes \citep{savage2012re}, outflows from patchy and bursty reconnection \citep{xue2020thermodynamical}, Rayleigh–Taylor instability (RTI; \citealt{guo2014rayleigh}), and a combination of RTI and Richtmyer–Meshkov instability (RMI; \citealt{shen2022origin}). In contrast to our observations, SADs are dark structures in EUV observations, whereas our observed downflows are brighter than the background. Furthermore, SADs occur in the supra-arcade fan, situated above certain heights in the corona, while our downflows can be traced from a height down to near the solar surface. Additionally, SADs are typically located above the flare arcades, whereas our downflows are located near the footpoints of magnetic field lines. Another similar phenomenon is coronal rain, which appears as cool and dense plasma falling down to the solar surface along loop legs due to thermal non-equilibrium and thermal instability (e.g., \citealt{antolin2022multi,chen2022coronal}). The falling materials normally move at speeds of ten to a few hundred km s$^{-1}$, typically observed in chromospheric lines and transition region lines with temperatures ranging from a few thousand to several hundred thousand Kelvin (e.g., \citealt{muller2005high,antolin2010coronal}). Some models show that coronal rain can be associated with magnetic reconnection \citep{li2018coronal,li2019repeated,li2020relation}. For example, \citet{li2018coronal} investigated the condensation of hot coronal plasma facilitated by interchange reconnection, involving the reconnection between higher-lying open structures and low-lying closed loops. This condensation leads to cool and dense coronal rain falling along the loops. However, in our event, the downflows can be clearly seen in high-temperature lines (211~\AA) and are less clear in low transition region lines (304~\AA) as shown in Fig. 4. Moreover, we do not observe a clear cooling process, hence we think that the mechanism of the downflows in our event is different from that of coronal rain.

\citet{innes2016analysis} investigated the impact of redirected filament materials hitting the solar surface. The redirected plasma fell at speeds of 230-450 km s$^{-1}$, causing UV and EUV intensity enhancements on the solar surface. Light curves in different passbands show a lag in the enhancement in high-temperature bands, indicating a heating process. In our event, however, the intensity enhancements in different passbands occur almost simultaneously (Fig. 5), indicating density enhancement or impulsive heating \citep{lionello2016can}.

The continuous ar-rf reconnection will cause footpoint drifting, as predicted by \citet{aulanier2019drifting}. This has been supported by some subsequent observations \citep{lorinvcik2019manifestations,zemanova2019observations,chen2019observational}. We also observed a continuous change of the locations of the downflows. As depicted by the red arrows in Fig. 3, the downflows initially appear in the middle of the box while most of them appear in the west afterwards, with a maximum drifting distance of about 110 Mm. Moreover, the downflows near the western footpoint also exhibit prominent drifting. The drifting motion of downflows likely correpond to the footpoint drifting in \citet{aulanier2019drifting}. There is also some less significant eastward drifting, which might be attributed to the magnetic configuration, where some of the overlying magnetic field lines are located progressively higher as the footpoint moves eastward. Consequently, a primary westward drifting is observed, accompanied by slight eastward drifting.
We also found that as the eruption progresses, the downflows can be observed at an increasingly larger height (see Fig. 4 (b1)-(b2)). This could be attributed to the increasing altitude of the reconnection sites, resulting in higher origins of the downflows.

After a rough examination of previous quiescent filament eruption events, we found that the occurrences of these downflows are not rare. Examples include SOL2013-09-29, SOL2023-05-06, SOL2023-06-12, SOL2023-06-25, and so on. Among these events, downflows consistently coincide with the hook-shaped structures of the flare ribbons, i.e., the footpoints of the MFR, suggesting that the filament reconnecting with overlying arcades is a common feature during filament eruptions. The resultant downflows may serve as markers for the footpoints of MFRs. This provides an efficient means to track MFR footpoints during eruptions, which might be important to investigate and forecast the propagation of interplanetary CMEs.

\section{Conclusion} \label{sec:conclusion}

In this study, we have reported a polar crown filament eruption, characterized by fast downflows below the filament. The unwinding motions at the legs of the filament reveal that the filament might be supported by a MFR. The downflows occur immediately after the onset of the eruption and persist for approximately 2 hours, at speeds of 92-144 km s$^{-1}$. They originate from the leading edge of the filament and there is no clear acceleration of the downflows. Interestingly, they are located at the opposite ends of the conjugate flare ribbons, which is similar to the typical distribution of the hooks of the QSL footprints of a MFR. This characteristic hints at an association of the downflows with the footpoints of the MFR. The downflows likely result from the continuous reconnection between the MFR and overlying magnetic field structures, known as the ar-rf reconnection in \citet{aulanier2019drifting}, and could be either reconnection outflows or reconnection-caused redirected filament materials. We also observed a driting of the locations of downflows near both footpoints, which likely correspond to the footpoint drifting predicted by \citet{aulanier2019drifting}. These downflows can serve as a potential indicator for the footpoints of MFRs, offering a method to track the footpoints of MFRs during eruptions.

\acknowledgments
This work is supported by the National Key R\&D Program of China (2022YFF0503800 and 2021YFA1600500) and the National Natural Science Foundations of China (12222306). D.Y.D was supported by the Beijing Natural Science Foundation (1244053) and National Postdoctoral Programs (GZC20230097, 2023M740112). SUTRI is a collaborative project conducted by the National Astronomical Observatories of CAS, Peking University, Tongji University, Xi'an Institute of Optics and Precision Mechanics of CAS and the Innovation Academy for Microsatellites of CAS. The CHASE mission is supported by China National Space Administration (CNSA). AIA is a payload onboard \emph{SDO}, a mission of NASA's Living With a Star Program. We thank Prof. Bin Chen and Dr. Sijie Yu for insightful discussions.

\bibliographystyle{aasjournal}
\bibliography{ref}

\clearpage 

\begin{figure}
    \centering
    \plotone{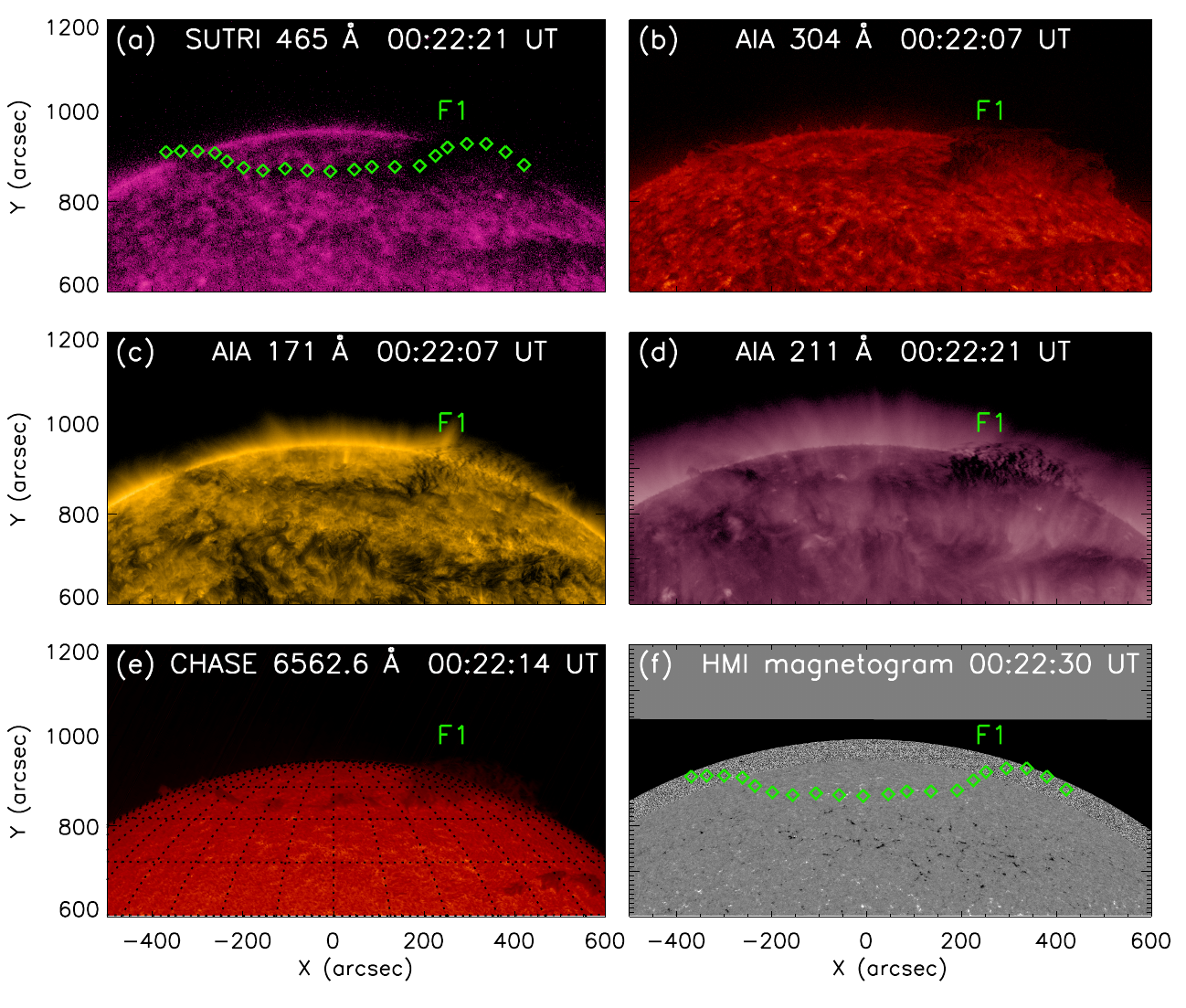}
  \caption{Pre-eruption observations of the polar crown filament. Panels (a)-(f) show the SUTRI 465~\AA, AIA 304~\AA, AIA 171~\AA, AIA 211~\AA, CHASE H$\alpha$, and HMI longitudinal magnetic field observations. The green diamonds denote the location of the polar crown filament. (An animation from 02:00 UT to 05:00 UT of this figure is available, which presents the eruption process of the polar crown filament eruption in AIA 304~\AA, 171~\AA, and 211~\AA\.)
    \label{fig:Fig.1}}
  \end{figure}
  
  \begin{figure}
    \centering
    \plotone{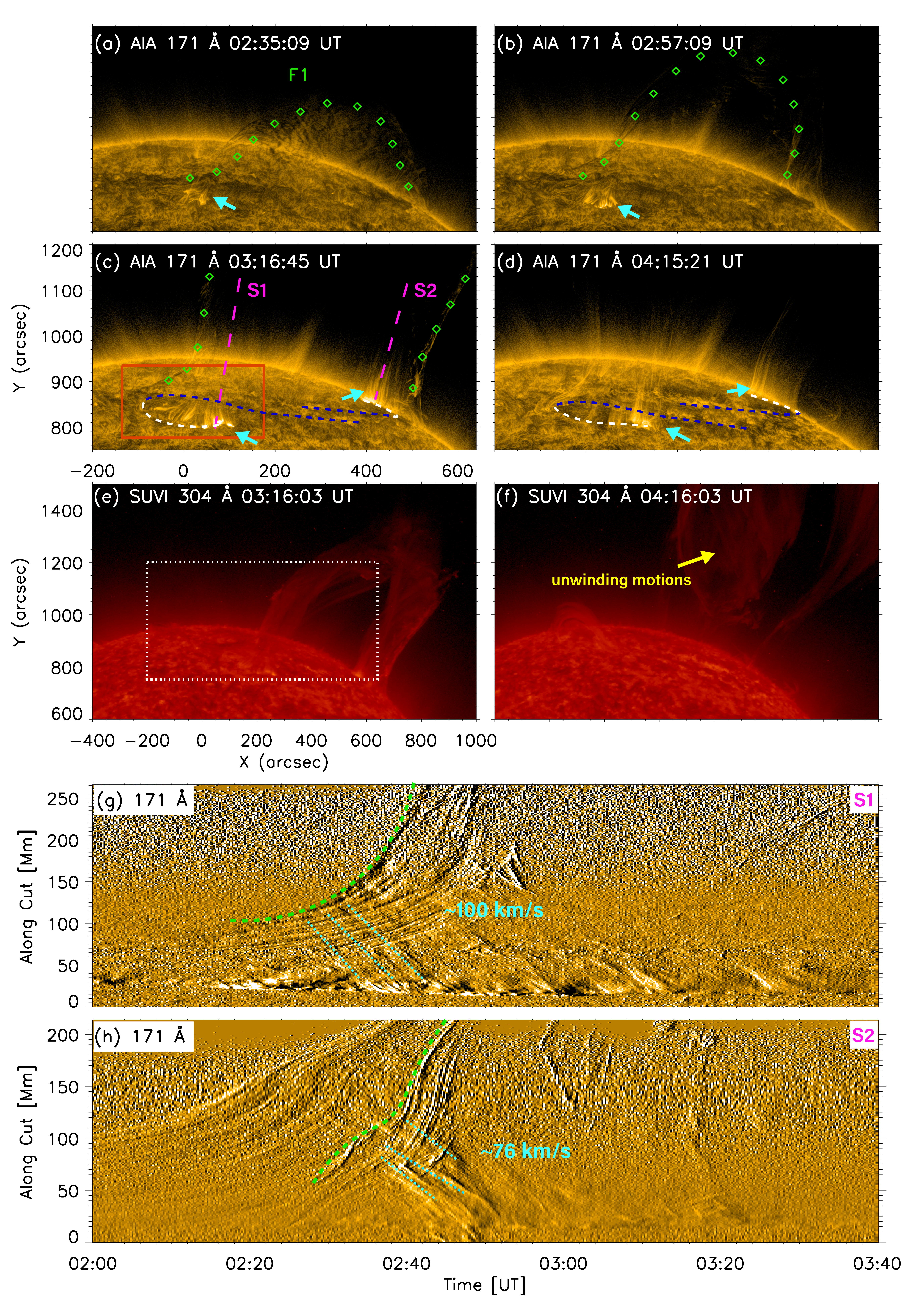}
  \caption{Overview of the filament eruption. Panels (a)-(d) show the eruption process in 171~\AA. The images are enhanced by using the MGN technique. The green diamonds and cyan arrows indicate the location of the polar crown filament and fast downflows, respectively. The blue dashed lines present the observed flare ribbons while the white dashed lines indicate the location where downflows occur. The pink dashed lines are selected to make time-distance diagrams shown in panels (g)-(h), which aim to show the trajectory of the filament and the origin of the downflows. Panels (e)-(f) display the SUVI 304~\AA\ observations, with the white dashed box in panel (e) denoting the FOV of panels (a)-(d). The yellow arrow in panel (f) indicates the observed unwinding motions at the legs of the filament.
  The green and cyan dashed lines in panels (e)-(f) display the filament and downflows, respectively. (An animation from 00:00 UT to 06:25 UT of this figure is available, which presents the eruption process of the polar crown filament eruption in GOES 304~\AA\.)
    \label{fig:Fig.1}}
  \end{figure}
  
  \begin{figure}
    \centering
    \plotone{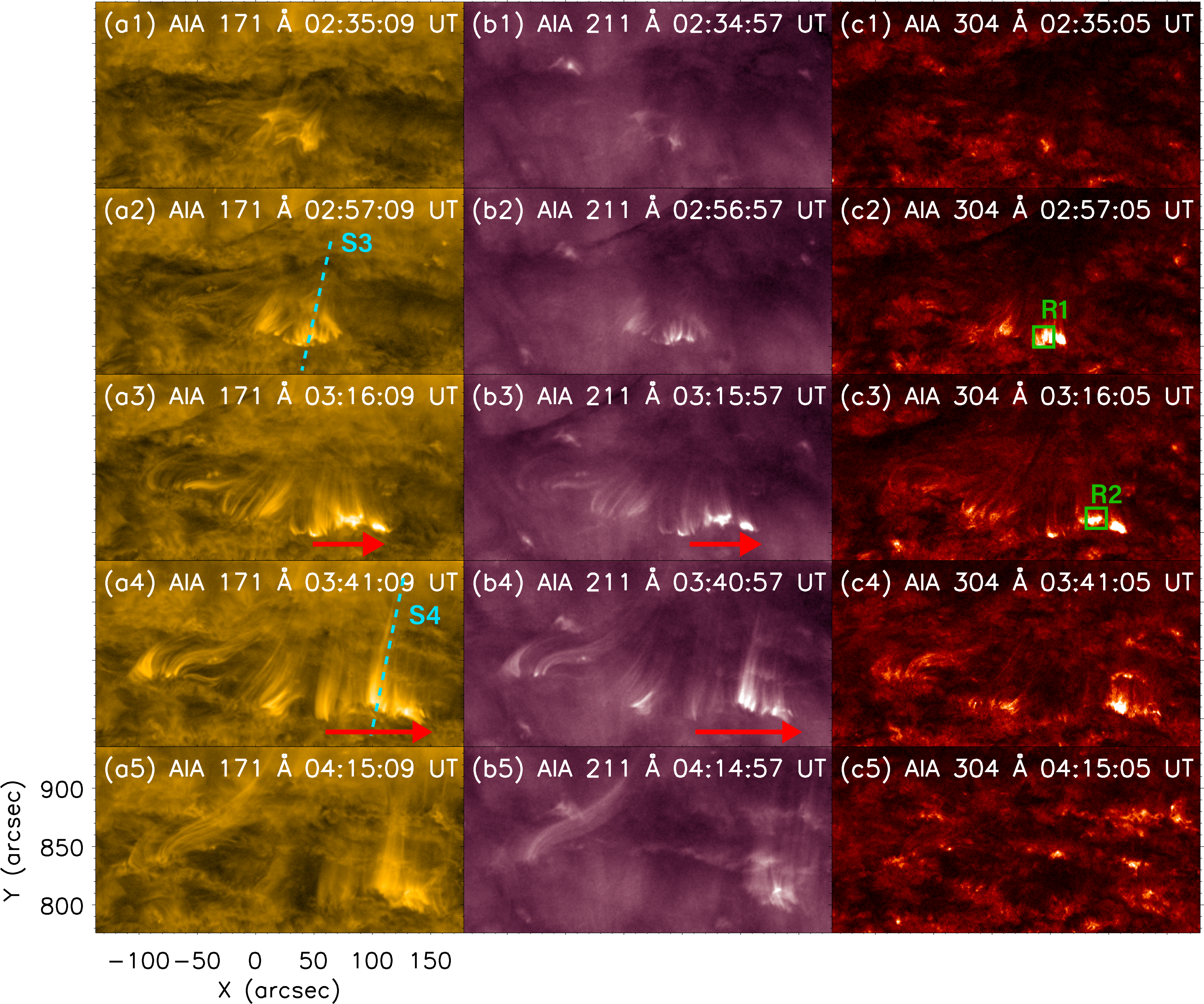}
  \caption{Zoom-in view of the fast downflows in 171~\AA\ (panels (a1)-(a5)), 211~\AA\ (panels (b1)-(b5)), and 304~\AA\ (panels (c1)-(c5)). The cyan dashed lines are chosen to make time-distance diagrams shown in Figure 4. The red arrows indicates the westward drifting of the locations of the downflows. The green boxes are selected to produce light curves, as depicted in Figure 5. (An animation from 02:00 UT to 05:00 UT of this figure is available, which presents the downflows in 171~\AA, 211~\AA, 304~\AA, and 131~\AA.)
    \label{fig:Fig.1}}
  \end{figure}
  
  \begin{figure}
    \centering
    \plotone{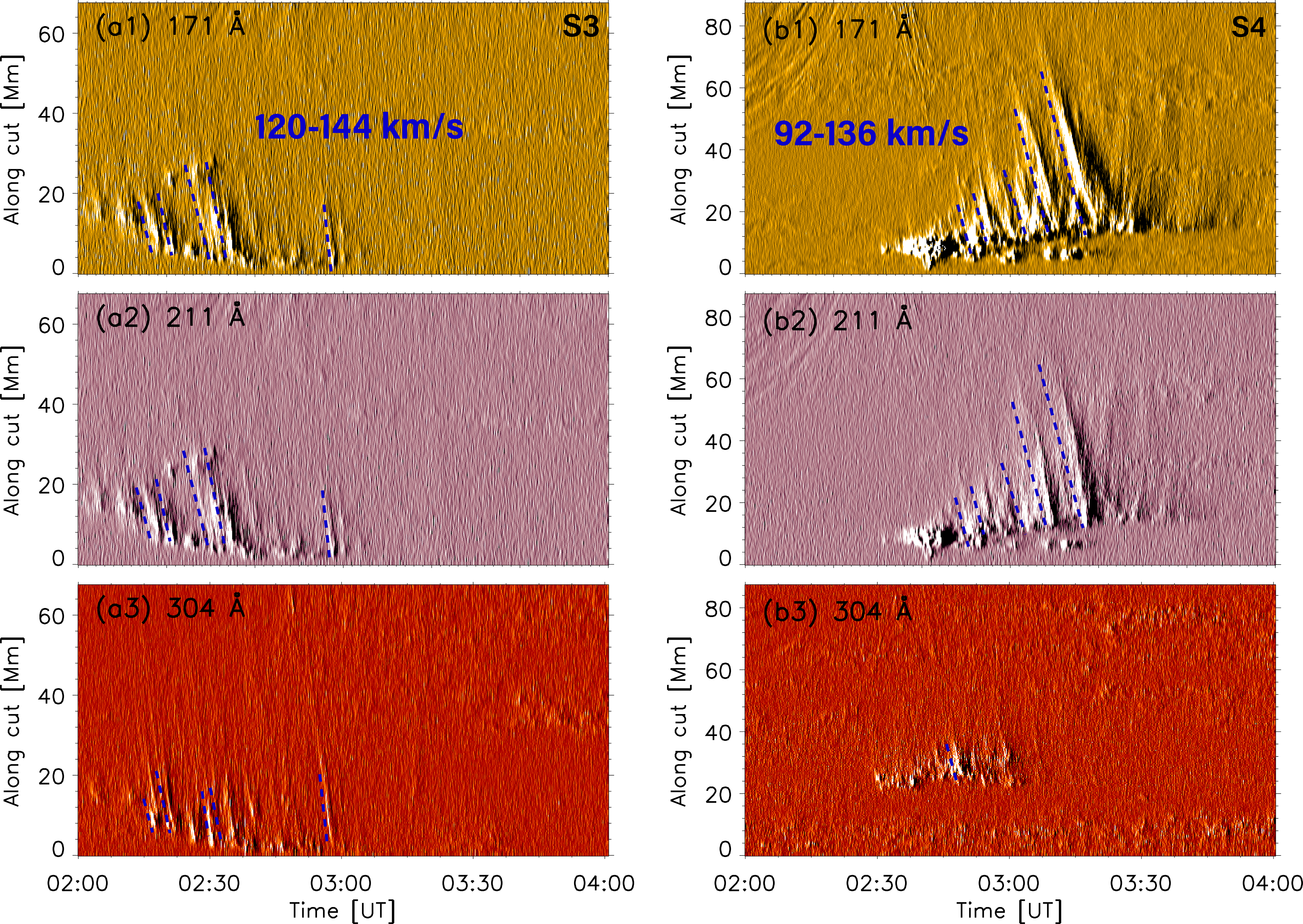}
  \caption{Time-distance diagrams in 171~\AA\ (panels (a1)-(b1)), 211~\AA\ (panels (a2)-(b2)), and 304~\AA\ (panels (a3)-(b3)), displaying the kinematic properties of the downflows. Panels (a1)-(a3) and (b1)-(b3) are derived from S3 and S4 in Figure 3, respectively.The blue dashed lines highlight several distinct downflows and the blue texts indicate the calculated POS speed of the downflows using linear fitting.
    \label{fig:Fig.1}}
  \end{figure}
  
  \begin{figure}
    \centering
    \plotone{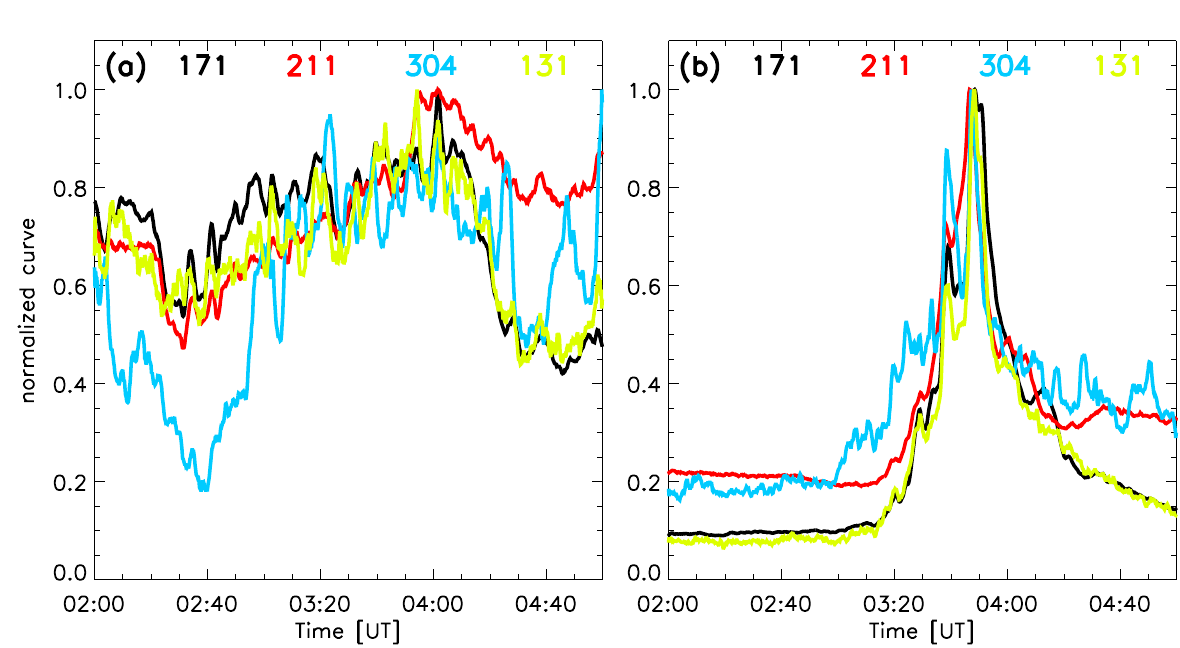}
  \caption{EUV response of the downflows near the solar surface. Panel (a) and (b) are light curves derived from R1 and R2 in Figure 3, respectively. The average intensities in the boxes are calculated in 171~\AA, 211~\AA, 304~\AA, and 131~\AA.
    \label{fig:Fig.1}}
  \end{figure}

\end{document}